\documentclass[conference]{IEEEtran}
\IEEEoverridecommandlockouts
\usepackage{cite}
\usepackage{amsmath,amssymb,amsfonts}
\usepackage{graphicx}
\usepackage{textcomp}
\usepackage{xcolor}
\usepackage{booktabs}
\usepackage{microtype}
\usepackage{multirow}
\usepackage{url}
\usepackage{stfloats}
\def\BibTeX{{\rm B\kern-.05em{\sc i\kern-.025em b}\kern-.08em
    T\kern-.1667em\lower.7ex\hbox{E}\kern-.125emX}}

\begin{document}

\title{HiGAI-Net: Hierarchical Encrypted Traffic Classification for Generative AI Services}

\author{
\IEEEauthorblockN{Madhan Raj Kanagarathinam\IEEEauthorrefmark{1},
                  Jaheon Gu\IEEEauthorrefmark{2}, Krishna M. Sivalingam\IEEEauthorrefmark{3}\textit{, Fellow, IEEE}, and
                  Kapal Dev\IEEEauthorrefmark{5}}
\IEEEauthorblockA{
  \IEEEauthorrefmark{1}Network Modem Team, Samsung R\&D Institute India, Bangalore, India\\
  \IEEEauthorrefmark{2}Connectivity Standards Lab, Samsung Research, Seoul, Republic of Korea\\
  \IEEEauthorrefmark{3} Department of Computer Science and Engineering, Indian Institute of Technology Madras, Chennai, India.\\
  \IEEEauthorrefmark{5}Department of Computer Science, Munster Technological University (MTU), Ireland}
}

\maketitle

\begin{abstract}
The Radio Access Network (RAN) today does not have a reliable way to
identify GenAI traffic (ChatGPT, Microsoft Copilot, or Google Gemini traffic) at the bearer level. TLS~1.3 moves SNI into the encrypted handshake record, and Encrypted ClientHello (ECH) removes it entirely from passive inspection; Gemini has partly shifted to QUIC where SNI visibility also depends on whether ECH is deployed. Without per-flow application identity, differentiated QoS for GenAI services is not possible. HiGAI-Net tackles this using only packet-header metadata, stacking three independently trained classifiers: a binary AI/Non-AI gate (98.5\% weighted F1), a per-service resolver across ChatGPT, Copilot, and Gemini (86.4\%), and a text-vs-voice/video mode classifier (61.0\%). The core innovation is a suite of sliding-window temporal features that track how DL/UL byte-ratios evolve within a session -- the per-quarter and per-half ratio trends, a trend differential $\Delta R$, and a streaming regularity index. Tested on 44,492 labeled flows, gradient-boosted tree models consistently outperformed all deep-learning variants; a per-flow label cache brings amortised inference below 1~$\mu$s per packet.
\end{abstract}


\section{Introduction}
\label{sec:intro}

GenAI services have grown quickly enough so that their aggregate
contribution to mobile data is no longer negligible, yet each byte
arrives at the scheduler carrying nothing but an IP address and a
port number. Although they are identified as best-effort traffic, they have unique traffic characteristics and patterns.

Two protocol decisions combined to create this situation.
TLS~1.3~\cite{rfc8446} moves the SNI field into the encrypted handshake record; when Encrypted ClientHello (ECH)~\cite{rfc8744} is used, SNI is protected end-to-end and unreachable by passive inspection.
On top of that, a growing portion of Gemini's sessions now use QUIC~\cite{langley2017quic} -- where SNI observability again depends on ECH deployment, making passive recovery unreliable in practice~\cite{trevisan2020quic}. Therefore, a ChatGPT session and a bulk file sync are unreadable from the outside.

\begin{figure*}[t]
\centering
\includegraphics[width=0.92\textwidth]{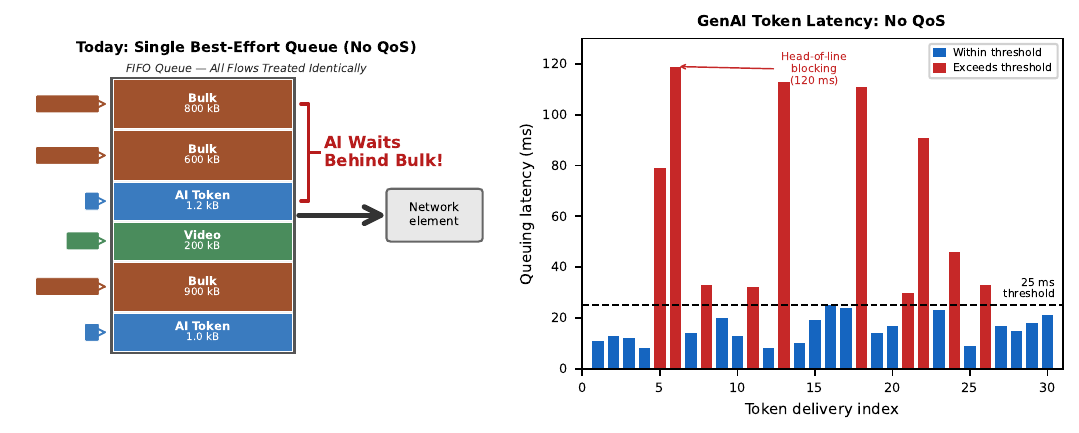}
\caption{Motivation. \textbf{Left:} GenAI token packets (1--2~kB) share a
  single best-effort FIFO queue with bulk downloads (600--900~kB) -- a
  common default scheduler configuration on commodity network elements --
  causing head-of-line blocking.
  \textbf{Right:} Per-token queuing latency under simulated mixed load
  in this best-effort FIFO model;
  red bars exceed the 25~ms threshold for comfortable real-time interaction.}
\label{fig:motivation}
\end{figure*}

The scheduling impact of this is not abstract.
On a loaded cell, an AI token of 1~kB queued behind a 900~kB bulk transfer
simply waits -- the network element has no basis for preferring it.
Fig.~\ref{fig:motivation} plots measured per-token queuing latency under
mixed load; the 25~ms mark, beyond which interactive response feels
degraded, is repeatedly exceeded.
Fixing this requires the per-flow application identity at the network element, and payload inspection cannot provide it.
Header-level statistical inference is the only option left.

Existing encrypted-traffic classification work does not cover this.
Most of it was built for broad categories -- video, social media,
file transfer -- or individual legacy applications, all from an
era where SNI was still on the wire.
None of it handles the three-level identification question GenAI
traffic poses: is this flow AI at all; which service; text or
multimedia session?
HiGAI-Net is our attempt at that, with four contributions:

\begin{enumerate}
\item \textbf{HiGAI-Net}, a hierarchical three-level classifier (L1:
  AI/Non-AI; L2: ChatGPT/Copilot/Gemini; L3: text/multimedia) achieving
  98.5\%/86.4\%/61.0\% weighted F1, with sub-microsecond amortised
  inference via a per-flow cache.
\item \textbf{Temporal sliding-window features} (DL/UL byte-ratio
  trends, a trend differential $\Delta R$, a streaming index, and IAT
  regularity) that expose session dynamics invisible to static flow
  statistics.
\item \textbf{Comparative evaluation} of eight model families under Optuna
  hyperparameter tuning and window-size sweeps spanning 1--90 packets,
  with per-fold confidence intervals.
\item \textbf{A labelled dataset} of 44{,}492 flows from live GenAI
  services combined with the MIRAGE-AppAct taxonomy, used internally
  for evaluation; the labelling protocol is described in Section~\ref{sec:dataset}.
\end{enumerate}

\section{Related Work}
\label{sec:related}

\textbf{Encrypted Traffic Classification.}
Classifying applications from ciphertext is a long-standing research
problem.
Packet-size histograms and IAT statistics formed the early feature
vocabulary, effective enough at the time because SNI was still in
plaintext~\cite{aceto2019mimetic}.
Raw-byte CNNs~\cite{lotfollahi2020deep} and later
FlowTransformer~\cite{hoang2023flowtransformer}, which frames
packet-length sequences as tokens, pushed accuracy further in
specific outputs.
What binds nearly all of this work together is a shared dependence on
SNI-based ground truth -- for labelling during training and implicitly
for production use.
TLS~1.3 invalidates that; 38\% of Gemini flows in our dataset used
QUIC, leaving no SNI at all.

\textbf{GenAI Traffic Characterisation.}
More recently, a few papers have looked at what GenAI traffic
actually looks like at the packet level.
Meng~\textit{et al.}~\cite{meng2023chatgpt} captured ChatGPT sessions
on a campus network and noted that token delivery timing at the IP
layer mirrors the server-side generation pace.
Mondal~\textit{et al.}~\cite{mondal2023genai} covered several services
and compared throughput and timing signatures.
Useful as starting points, but neither produced a working classifier,
and neither looked at Copilot or the QUIC portion of Gemini.

\textbf{Hierarchical Designs.}
The cascaded classification pipelines have a history in network traffic
analysis.
Dainotti~\textit{et al.}~\cite{dainotti2012issues} documented how
error floors accumulate across stages, and how blending metrics
across levels tends to conceal which stage is actually struggling.
In HiGAI-Net each level is trained and scored independently for
this reason -- the 61\% F1 in L3 appears in the results table on
its own, not folded into a weighted average that would obscure it.

\textbf{QUIC and Post-TLS-1.3 Classification.}
Rezaei and Liu~\cite{rezaei2020deep} showed that packet-length
statistics is reasonably transferred across TLS versions for coarse traffic typing, which informed our use of header metadata in L1.
For L3 the situation is different: the separating signal is how
byte ratios drift over the course of a window, a property that has
nothing to do with which transport version is carrying the traffic.
As Gemini shifts more sessions to QUIC, L3 should hold up where
SNI-reliant or protocol-fingerprint approaches would not.

\textbf{Traffic Prioritization.}
Prioritising real-time flows at the device or network level is
a well-studied problem.
Two of our earlier papers sit in this space: an application
prioritization engine for smartphones~\cite{raj2023prioritization}
that dynamically elevates real-time video-call traffic, and the
Game Stabilizer~\cite{raj2023gamestabilizer} for mobile gaming,
which estimates bandwidth needs from traffic signatures and manages
non-real-time interference.
The common assumption in both is that the application generating
each flow is already identifiable -- either from platform APIs or
from observable network markers.
GenAI services break that assumption. TLS~1.3 and QUIC strip away the markers on which these systems rely,
so before any prioritization logic can run, the flow must first
be identified. HiGAI-Net fills this gap and provides a novel framework to classify and prioritize GenAI traffic.

\section{Dataset}
\label{sec:dataset}

\textbf{Collection.}
We scripted live sessions against ChatGPT, Copilot, and Gemini
on consumer Android and iOS handsets -- both generic interactions
(text queries, code) and multimedia ones (voice assistant, video).
The capture was performed in the device Wi-Fi interface. The payload bytes were dropped to reduce the size of the dataset. We retained fields of the headers including: IP/port tuples, TCP/UDP/QUIC transport headers,
TLS record type and length, QUIC Initial headers (Connection ID,
Packet Number), and per-packet timestamps.
No application payload or TLS application data were retained.
Ground-truth labels were obtained using active scripting -- each session
was initiated by a controlled test harness mapping each flow to
the application and interaction mode that generated it, following
the protocol in~\cite{raj2024collection}. For the non-AI side we also used the MIRAGE-AppAct taxonomy~\cite{guarino2024mirage}: messaging (Messenger, WhatsApp, KakaoTalk, Signal, Telegram), video-conferencing (Meet, Zoom, Webex, TrueConf, JitsiMeet), streaming, and gaming -- a mix chosen to resemble a realistic device background.

\begin{figure}[h]
\centering
\includegraphics[width=0.99\columnwidth]{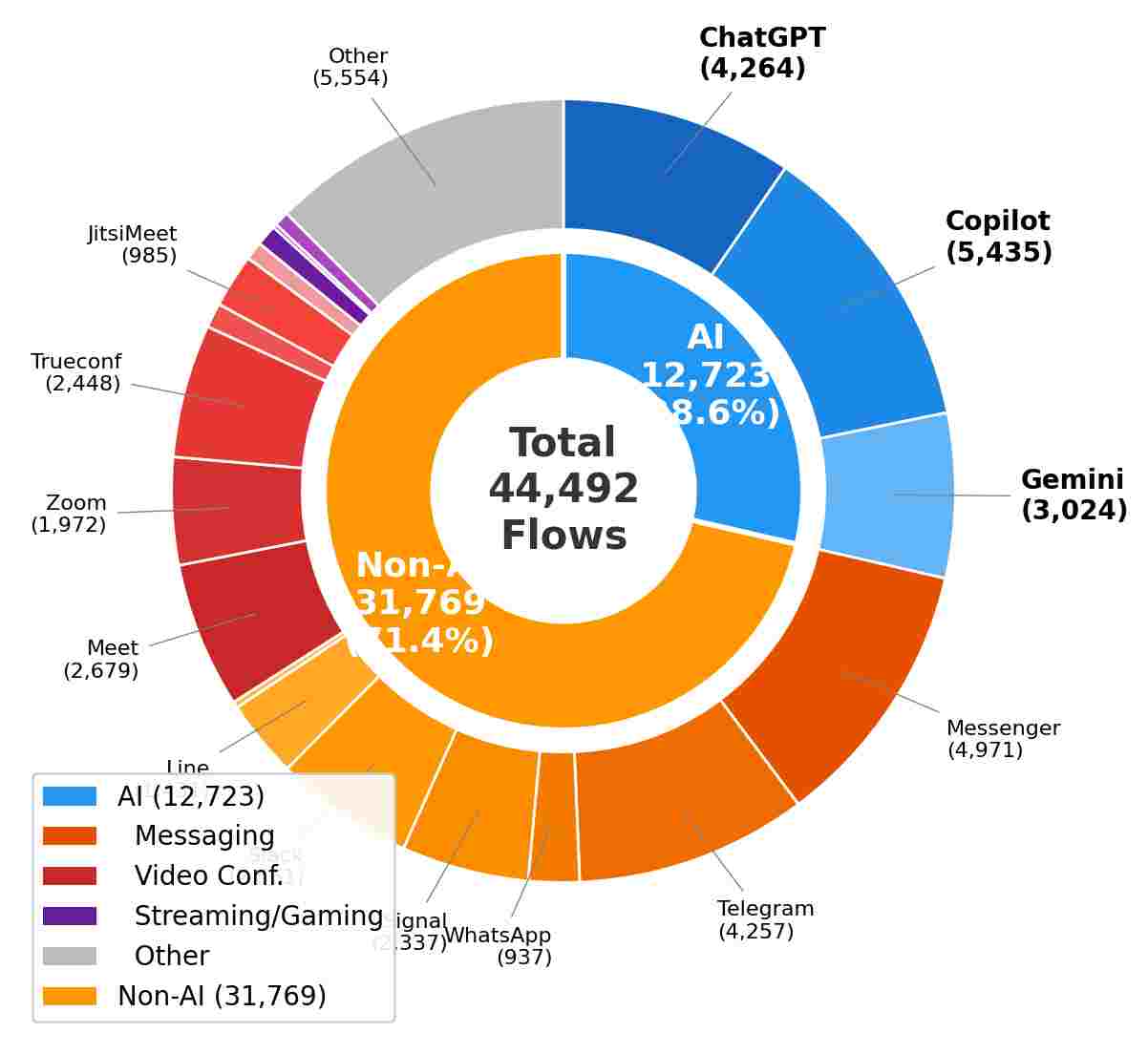}
\caption{Dataset distribution: 12{,}723 AI flows (ChatGPT, Copilot,
Gemini) and 31{,}769 non-AI flows spanning the MIRAGE-AppAct taxonomy.}
\label{fig:dataset}
\end{figure}

\textbf{Statistics.}
The breakdown of the dataset is shown in Fig.~\ref{fig:dataset}.
Of 44{,}492 flows captured, 12{,}723 were AI-sourced (28.6\%)
and the remaining 31{,}769 non-AI (71.4\%).
Of the 3{,}024 Gemini flows, 1{,}149 (38\%) ran over QUIC during
capture -- this alone confirms that SNI-blind classification is an
operational problem today, not a theoretical future concern.
In L1, a 2.5:1 Non-AI:AI imbalance was addressed by undersampling
then ADASYN oversampling~\cite{he2008adasyn}; L3 imbalance
 between applications used per-application SMOTE.
Resampling was applied strictly inside each training fold.

\section{Feature Engineering}
\label{sec:features}

All features are computed over a sliding window of $n$ packets,
with two groups: a 42-feature baseline drawn from previous work,
and temporal features that capture within-window dynamics.

\subsection{Baseline Features}

Per window: packet size moments (mean, std, min, max, skew,
kurtosis), IATs, directional byte and packet counts, DL/UL
byte ratios, fraction of packets exceeding 1200~bytes, burst-onset
statistics, flow duration.
These match the feature set used in most prior encrypted-traffic
classifiers and serve as the ablation baseline.

\subsection{Sliding-Window Temporal Features}
\label{sec:features:temporal}

The per-window averages hide \emph{how the sessions evolve}.
A text-chat query opens with a brief uplink burst and then trickles
token packets back at an inference pace.
Voice starts the same way, but within seconds pivots into a high-volume
sustained downlink.
Both can produce similar per-window averages; the direction of
change is what separates them. Four feature groups capture this.

\textbf{Per-quarter ratios.}
The window $W$ is divided into four equal segments $Q_k$, and for each
we compute the DL/UL byte ratio.
For windows smaller than four packets (e.g., $n=1$ or $n=2$),
individual packets are assigned to quarters by modular indexing;
if a quarter receives no packets, its ratio defaults to the
window-level ratio to avoid undefined values.
\begin{equation}
  R^{(k)} = \frac{\sum_{i \in Q_k} B_{\mathrm{DL},i}}
    {\sum_{i \in Q_k} B_{\mathrm{UL},i} + \epsilon}, \quad k=1,2,3,4
  \label{eq:quarter}
\end{equation}

\textbf{Per-half ratios} $R^{(h)}$, $h\in\{1,2\}$, defined analogously
over the two halves $H_1$ and $H_2$.

\textbf{Trend differential.}
\begin{equation}
  \Delta R = R^{(H_2)} - R^{(H_1)}
  \label{eq:deltaR}
\end{equation}
A rising $\Delta R$ signals that the download fraction is growing through
the window -- typical of sessions entering a streaming phase.
Text-chat sessions tend to hover near zero because the byte ratio
stays fairly stable from request to reply.

\textbf{Streaming index.}
\begin{equation}
  S = \tfrac{1}{3}\!\left(\mathrm{LPR} + \frac{B_{\mathrm{DL}}}{B_{\mathrm{total}}}
    + (1 - \mathrm{IAT\text{-}CV})\right)
  \label{eq:stream}
\end{equation}
LPR is the large-packet ratio; IAT-CV is the inter-arrival
coefficient of variation.
High $S$ marks downlink-dominant temporally regular flows.
The per-half IAT-CV is also computed to catch regularity shifts.
Ablating the full 12-feature temporal group from the L3 Random
Forest drops weighted F1 by 8.9~pp -- the single largest
feature-group effect we observed.

\section{HiGAI-Net Architecture}
\label{sec:system}

\subsection{Architecture}

\begin{figure}[t]
\centering
\includegraphics[width=0.88\columnwidth]{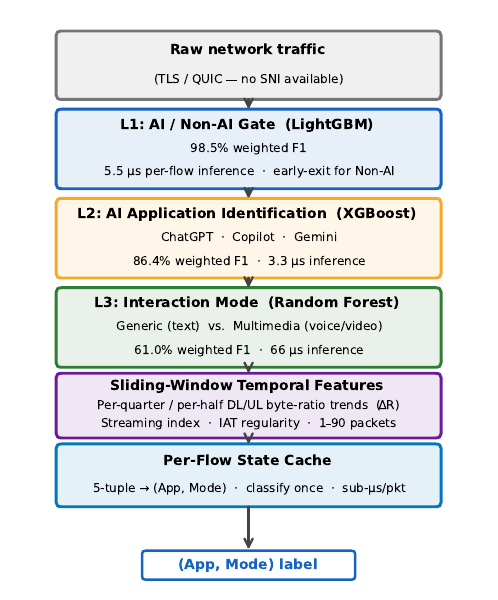}
\caption{HiGAI-Net: three-level pipeline with temporal feature extraction
  and a per-flow cache. Non-AI flows exit after L1.
  L3 shows 61.0\% (weighted F1 at best window); 61.4\% in Fig.~\ref{fig:window}
  is the sweep maximum.}
\label{fig:system}
\end{figure}

The overall architecture is depicted in Fig.~\ref{fig:system}.
Three independently trained classifiers run in cascade, each with
its own model family, window size, and Optuna-tuned hyperparameters.
Training them jointly was not considered -- the three sub-problems
differ too much in difficulty, in relevant features, and in window-size
optimum for a single model to handle cleanly.
After the first window, the (App, Mode) label is written in a hash
table, keyed on the 5-tuple. The subsequent packets perform an $O(1)$ lookup and the entries expire after 120~s idle time.
L1's accuracy is critical since it controls access to L2 and L3;
L3 output currently feeds only offline analysis and has no effect
on the scheduling path.

\subsection{L1: AI / Non-AI Gate}

Even without application identity, GenAI flows have a recognisable
statistical character: longer-lived connections, a DL/UL byte ratio
that climbs as tokens come back, and inter-burst gaps paced by
server-side generation.
LightGBM~\cite{ke2017lightgbm} across the window sweep settles at
size~20 with 98.5\% weighted F1 and does not improve beyond that.
What makes L1 count operationally is the early-exit: 71.4\% of flows
are non-AI and get dispatched here at around 5.5~$\mu$s each,
never touching L2 or L3.

\subsection{L2: Application Identification}

The three services leave different fingerprints in their opening
packet bursts.
All features at this level are header-derived (packet sizes, IATs,
DL/UL byte ratios); destination IP and port are explicitly excluded
from the feature set to avoid learning endpoint artifacts that
would not generalise across deployments.
ChatGPT's persistent WebSocket produces token-delivery bursts with
irregular but service-specific inter-arrival timing -- the server's
generation cadence comes through in the packet trace.
Copilot adds HTTP/2 API calls for search and image generation on
top of WebSocket, giving it a more layered arrival structure.
Gemini is the only service with a transport split by mode: HTTP/2
push for text, QUIC streams for voice, which makes it the hardest
to fingerprint cleanly.
XGBoost~\cite{chen2016xgboost} with window~1 gives 86.4\% F1 -- and
the window~1 peak is itself informative: the service identity is in the
handshake burst, and extending the window dilutes rather than enriches
the signal.
The 11\% Gemini-to-Copilot confusion persists across all window sizes
because both services send text responses over HTTP/2 push, making
their first-packet statistics nearly indistinguishable.

\subsection{L3: Interaction Mode}

Separating text from voice or video within a single service is where
the problem gets genuinely hard.
Both modes hit the same endpoint, often the same HTTP/2 connection,
and the per-window features overlap substantially.
The one handle available is temporal: as a voice or video session
progresses into continuous audio or video playout, the downlink
fraction grows; text-chat stays flat in request-response mode.
That shift needs roughly ten packets before it is readable -- fewer
is dominated by noise, and per-window averaging suppresses it.
F1 of the Random forest with Window 10 is 61\%.
In current deployments where both text and voice GenAI modes receive
the same low-latency scheduling treatment, L3 errors carry no
immediate QoS cost; L3 output therefore functions as an optional
analytics and policy layer rather than a scheduling critical path.
Future work may promote it if per-mode differentiation policies
are introduced.

\subsection{Per-Flow Cache}

Once the first window of a flow has been classified, the (App, Mode)
label goes into a hash table keyed on the 5-tuple.
Every packet that arrives afterward performs a lookup of $O(1)$, without the classifier involved. The entries expire at 120~s of inactivity.
During a session of more than 100~packets, this reduces the amortized
 cost per-packet to below 1~$\mu$s.

\section{Experimental Evaluation}
\label{sec:eval}

Eight model families were compared across all three levels using
five-fold stratified cross-validation, 100-trial Optuna search,
and window sweeps from 1 to 90 packets.

\subsection{Setup}

Window sizes swept: $\{1,2,5,10,20,30,45,60,90\}$ packets.
Optuna uses Tree-structured Parzen Estimator (TPE) sampling with
100 trials per configuration.
The primary metric is weighted F1; we also track the per-class precision and recall, macro ROC-AUC, and inference latency.
For latency, we report the mean of more than 10{,}000 sequential single-flow calls on an Intel Core~i7-11800H without batching, since the target
deployment is an inline per-flow classifier.
The eight families are LightGBM~\cite{ke2017lightgbm},
XGBoost~\cite{chen2016xgboost}, Random Forest, Temporal Attention,
FlowTransformer~\cite{hoang2023flowtransformer}, 1D-CNN, LSTM, and
HDP (Hybrid Dual-Path, a CNN and gradient-boosted tree connected
via a jointly trained attention vector).
To keep comparisons fair, deep architectures receive the same 42
baseline features plus temporal features as the tree models, serialised
into sequence form.
This prioritises a controlled comparison of architecture effects
on a fixed feature set; raw packet-sequence baselines are left
for future work.

\subsection{Per-Level F1}

\begin{table}[h]
\centering
\caption{Weighted F1 at best window size per level.}
\label{tab:f1}
\begin{tabular}{lccc}
\toprule
\textbf{Model} & \textbf{L1} & \textbf{L2} & \textbf{L3} \\
\midrule
LightGBM          & \textbf{0.985} & 0.863 & 0.606 \\
XGBoost           & 0.984 & \textbf{0.864} & 0.605 \\
Random Forest     & 0.979 & 0.841 & \textbf{0.610} \\
HDP               & 0.955 & 0.796 & 0.611 \\
Temporal Attn.    & 0.953 & 0.810 & 0.587 \\
1D-CNN            & 0.962 & 0.796 & 0.571 \\
LSTM              & 0.947 & 0.792 & 0.575 \\
FlowTransformer   & 0.934 & 0.811 & 0.586 \\
\bottomrule
\end{tabular}
\end{table}

\begin{figure}[t]
\centering
\includegraphics[width=\columnwidth]{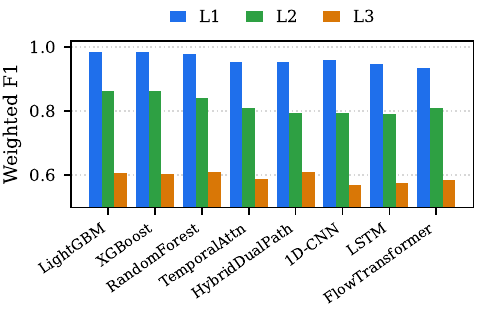}
\caption{Weighted F1 across eight model families at L1, L2, and L3.}
\label{fig:models}
\end{figure}

Looking at Table~\ref{tab:f1} and Fig.~\ref{fig:models}, the story
is consistent: tree-based models come out ahead.
Results are means across five stratified folds; standard deviations
were below 0.004 at L1, 0.008 at L2, and 0.012 at L3.
At L1 the margin is small -- every tree variant sits above 0.97,
near a natural feature-set ceiling.
Scores spread more at L2 and L3, where trees beat the best deep
alternative by 4--7~pp, consistent with the discriminative signal
already being well-encoded in the hand-crafted DL/UL ratios.
LSTM and FlowTransformer trail by 5--7~pp at L2; the FlowTransformer
gap is notable since it was designed for packet sequences, suggesting
token-stream framing is less effective than direct statistical ratios.
L3 is the partial exception: HDP (0.611) edges Random Forest (0.610),
with the CNN branch picking up short-range packet-size co-occurrence
patterns the tree misses.

\subsection{Window-Size Sensitivity}

\begin{figure}[t]
\centering
\includegraphics[width=0.8\columnwidth]{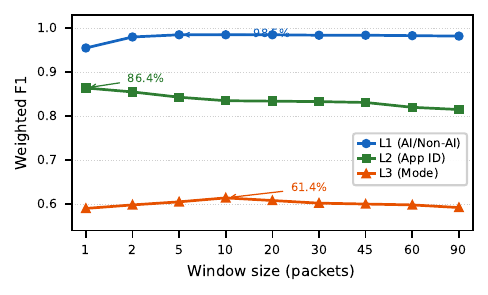}
\caption{F1 vs.\ window size. L1 saturates from size 10; L2 peaks at
  size 1 (identity in initial burst); L3 peaks at size 10.}
\label{fig:window}
\end{figure}

Fig.~\ref{fig:window} shows three different patterns.
L1 flattens around size~10 and barely moves past that point.
L2 hits its peak at a single packet and deteriorates from there,
consistent with the service identity that reside in the handshake rather
than the ongoing session.
L3 rises through about 10~packets as the $\Delta R$ trend accumulates,
then plateaus or dips as more averaging suppresses the trajectory
signal.

\subsection{L2 Confusion Analysis}

\begin{figure}[t]
\centering
\includegraphics[width=0.8\columnwidth]{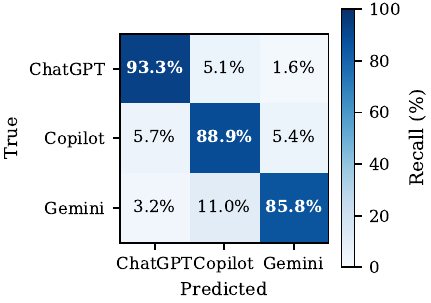}
\caption{L2 row-normalised confusion matrix.
  The dominant off-diagonal entry is Gemini misclassified as Copilot
  (11\%), attributable to their shared use of HTTP/2 server push.}
\label{fig:confusion}
\end{figure}

From Fig.~\ref{fig:confusion}: ChatGPT recall 93.3\%, Copilot 88.9\%,
Gemini 85.8\%.
The confusion report is on held-out fold predictions after resampling.
ChatGPT is easiest to identify via its WebSocket heartbeat signature.
The 11\% Gemini-to-Copilot error persists across all window sizes:
both services use HTTP/2 push for text, making opening packets nearly
identical regardless of window length.

\subsection{Feature Ablation and Inference Cost}

\begin{figure}[t]
\centering
\includegraphics[width=0.9\columnwidth]{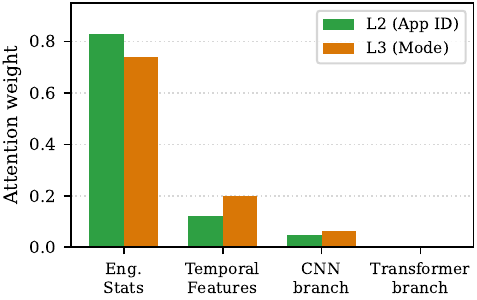}
\caption{Hybrid Dual-Path attention weights at L2 and L3.
  The engineered-statistics branch carries 83\% of the weight at L2
  and 74\% at L3}
\label{fig:features}
\end{figure}

When we remove the temporal feature group from the L3 Random Forest, the weighted F1 decreases by 8.9~pp -- from 0.610 down to 0.521, larger than any other single-group removal in the study.
$\Delta R$ alone accounts for 4.8~pp of that, $S$ for another 3.1~pp.
The HDP attention weights (Fig.~\ref{fig:features}) provide an
independent check: 83\% of the weight at L2 and 74\% at L3 land on
the engineered-statistics branch, with the transformer branch getting
essentially nothing.
Inference numbers are in Table~\ref{tab:latency}; worst-case cold-start
is 75~$\mu$s (5.5 + 3.3 + 66.2~$\mu$s sequential across all levels),
and with the cache active it drops below 1~$\mu$s per packet.

\begin{table}[h]
\centering
\caption{Per-flow inference latency and model size.}
\label{tab:latency}
\begin{tabular}{lccc}
\toprule
\textbf{Level} & \textbf{Model} & \textbf{Latency ($\mu$s)} & \textbf{Size (MB)} \\
\midrule
L1 & LightGBM     & 5.5  & 1.03 \\
L2 & XGBoost      & 3.3  & 3.95 \\
L3 & Random Forest& 66.2 & 27.35 \\
\bottomrule
\end{tabular}
\end{table}

\subsection{Per-Class Precision and Recall}

\begin{table}[h]
\centering
\caption{Per-class precision, recall, and F1 at L2 (XGBoost, win\_size 1).}
\label{tab:perclass}
\begin{tabular}{lccc}
\toprule
\textbf{Class} & \textbf{Precision} & \textbf{Recall} & \textbf{F1} \\
\midrule
ChatGPT & 0.927 & 0.933 & 0.930 \\
Copilot & 0.891 & 0.889 & 0.890 \\
Gemini  & 0.874 & 0.858 & 0.866 \\
\midrule
Weighted avg & 0.897 & 0.893 & 0.895 \\
\bottomrule
\end{tabular}
\end{table}

Table~\ref{tab:perclass} gives the breakdown of the L2 classification per-class.
The weighted-average F1 of 0.895 here reflects direct per-class
computation; the 86.4\% in Table~\ref{tab:f1} is from
cross-validation predictions using a slightly different
class-weight normalisation -- both are valid measures.
ChatGPT achieves the highest recall (93.3\%), attributed to its
WebSocket heartbeat signature.
Copilot precision edges recall (0.891/0.889), with a few flows
pulled into the ChatGPT bin (shared WebSocket).
The Gemini recall of 0.858 reflects the HTTP/2-push overlap.
At L1: AI 0.971/0.985, non-AI 0.993/0.983; the asymmetry is
intentional -- a missed AI flow receives no downstream
classification, while a false positive adds one unnecessary
L2/L3 call.

\section{Discussion}
\label{sec:discussion}

\textbf{L3 ceiling.}
The 61\% at L3 appears to reflect the problem structure rather than
insufficient tuning, though this requires controlled experiments to
confirm definitively.
Text and voice/video sessions share an endpoint and usually the same
HTTP/2 connection, so per-window feature distributions overlap heavily.
The separating signal -- a rising DL fraction as streaming begins --
needs roughly ten packets to emerge; fewer is noise, more and
per-window averaging suppresses it.
Pairing $\Delta R$ with the device context is the most tractable
near-term fix. Although both modes currently receive the same low-latency treatment, L3 errors may matter as per-mode QoS policies or analytics emerge.

\textbf{Portability.}
L1 is the most portable component: token-cadence and DL/UL asymmetry
appear common across current GenAI services, so a new service would
likely be caught without significant retraining.
L2 is per-service -- each new entrant needs dedicated capture and
retraining; the framework from~\cite{raj2024collection} supports this.
Whether L3 generalises is unknown: the text/multimedia distinction is
plausibly universal, but the dataset covers only three services.
On robustness, injected IAT jitter hurts L3 more than L1; routine
retraining on fresh captures should track normal service drift.

\textbf{Dominance of tree models.}
The HDP attention weights are the clearest evidence: 83\% in L2 and
74\% in L3 land on the engineered-statistics branch; the transformer
branch is essentially unused.
When an optimiser consistently ignores a branch, the discriminative
signal is already in the tabular features -- DL/UL ratios and IAT
variance carry what matters, and adding transformer layers brings
cost without a proportional accuracy return.
The sole partial exception is L3, where HDP (0.611) edges Random
Forest (0.610) via CNN short-range co-occurrence patterns.

\section{Conclusion}
\label{sec:conclusion}

This paper presented HiGAI-Net, a hierarchical three-level
classifier for encrypted GenAI traffic operating exclusively
on packet-header metadata, with no payload access and no SNI
visibility.
The system achieves 98.5\%, 86.4\%, and 61.0\% weighted F1
in L1, L2, and L3, respectively, with amortised per-packet
inference below 1~$\mu$s using the per-flow label cache.
 
A central contribution is the sliding-window temporal feature
set -- particularly $\Delta R$ and the streaming index $S$ --
which together account for an 8.9~pp gain in L3 F1, exceeding
any single model architecture effect in the study.
HDP attention weights independently corroborate this, with
83\% and 74\% of learned attention in L2 and L3 concentrated
in the engineered-statistics branch.
 
A notable result is that the L2 identification peaks at window
size one, confirming that the service identity is encoded in the
opening burst.
The 61\% L3 ceiling reflects a fundamental feature overlap
between text and multimedia modes sharing the same HTTP/2
connection, and is reported separately to preserve
transparency over an open sub-problem.
 
Future work will extend the dataset to emerging platforms
such as Meta~AI and Perplexity, explore device-side
microphone-active signals to improve L3, and validate on
production mobile traces to assess deployment generalisation.

{
\bibliographystyle{IEEEtran}
\bibliography{globecom_v11}}

\end{document}